\documentclass[letter]{aa} 
\usepackage{graphicx}
\usepackage{amsmath}
\usepackage[dvips,usenames]{color}
\usepackage{natbib}
\usepackage{txfonts}
\newcommand{\realfigure}[3]{
  \begin{figure}
    \centering
    \resizebox{.95 \hsize}{!}{\includegraphics{#1}}
      \caption{#2}\label{#3}
    \end{figure}}

\newcommand{\uks}{UKS\,2323-326}

\newcommand{\feh}{{\rm [Fe/H]}}

\newcommand{\ebv}{\mbox{$E_{B\!-\!V}$}}

\newcommand{\evh}[1]{{#1}}

\newcommand{\referee}[1]{{#1}}

\begin{document}
   \title{Resolving Stellar Populations outside the Local Group: 
MAD observations of \uks
   %
     \thanks{
Based on observations collected at 
the European Southern Observatory, 
Chile, as part of MAD Guaranteed Time Observations,
on ESO archival observations
(Programme 71.D-0560), and on 
NASA/ESA Hubble Space Telescope observations 
(Proposal ID 8192).}}

\author{M. Gullieuszik
          \inst{1}\and
          L. Greggio
          \inst{1}\and
          E. V. Held
          \inst{1}\and
          A. Moretti
          \inst{1}\and
          C. Arcidiacono
          \inst{1}\and
          P. Bagnara
          \inst{1}\and
          A. Baruffolo
          \inst{1}\and
          E. Diolaiti
          \inst{2}\and
          R. Falomo
          \inst{1}\and
          J. Farinato
          \inst{1}\and
          M. Lombini
          \inst{2}\and
          R. Ragazzoni
          \inst{1}\and
          R. Brast
          \inst{3}\and
          R. Donaldson
          \inst{3}\and
          J. Kolb
          \inst{3}\and
          E. Marchetti
          \inst{3}\and
          S. Tordo
          \inst{3}
          }
   \offprints{marco.gullieuszik@oapd.inaf.it}
 \institute{Osservatorio Astronomico di Padova, INAF,
              vicolo dell'Osservatorio 5, 35122 Padova, Italy
         \and
             Osservatorio Astronomico di Bologna, INAF,
             via Ranzani 1, 40127 Bologna, Italy
         \and
          European Southern Observatory, 
          Karl-Schwarzschild-Strasse 2, 85748 Garching, Germany 
}

   \date{Received ... ; accepted ...}

 
  \abstract
  {} 
  {We present a study aimed at deriving constraints on star formation
    at intermediate ages from the evolved stellar populations in the
    dwarf irregular galaxy \uks.  These observations were also
    intended to demonstrate the scientific capabilities of the
    multi-conjugated adaptive optics demonstrator (MAD) implemented at
    the ESO Very Large Telescope as a test-bench of adaptive optics
    (AO) techniques.}  {We perform accurate, deep photometry of the
    field using $J$ and $K_s$ band AO images of the central region of
    the galaxy.}
{The near-infrared (IR) colour-magnitude diagrams clearly show the
  sequences of asymptotic giant branch (AGB) stars, red supergiants,
  and red giant branch (RGB) stars down to $\sim$1 mag below the RGB
  tip. Optical--near-IR diagrams, obtained by combining our data with
  Hubble Space Telescope observations, provide the best separation of stars in the
  various evolutionary stages.  The counts of AGB stars brighter than
  the RGB tip allow us to estimate the star formation at intermediate
  ages.  Assuming a Salpeter initial mass function, we find that the
  star formation episode at intermediate ages produced $\sim 6\times
  10 ^5$ $M_\odot$ of stars in the observed region.  }
   {}
 

   \maketitle
%

\section{Introduction}

The study of the resolved stellar populations in external galaxies has
developed greatly in the last decade to become arguably the most
accurate tool to investigate star formation history in stellar
systems.  However, with standard instrumentation at ground-based
telescopes this study is limited to the nearest galaxies, due to the
severe crowding of stars.  High-precision photometry for the most
distant galaxies in the Local Group (LG) and beyond can be obtained
only with the Hubble Space Telescope (HST).
  
New opportunities in this field are foreseen with the realisation of
imagers equipped with adaptive optics (AO), on the largest aperture
telescopes. The use of AO systems is mandatory for the future larger
($>$ 10m ) telescopes, but it can also significantly improve the
performances of telescopes already in operation.  A relevant example
is given by the multi-conjugated adaptive optics demonstrator (MAD)
recently developed by ESO \evh{(see next section)} that allows us to
test AO capabilities for stellar photometry on the sky.
 
In this context, as part of a Guaranteed Time Observations program, we
obtained MAD near-infrared (IR) images of the dwarf irregular galaxy
\uks \referee{ (UGCA\,438)}.  We chose this galaxy from a list of
targets selected according to various criteria: favourable position on
the sky with respect to the availability of stars to perform the AO
correction; \referee{low Galactic latitude ($b=-70\fdg9$), to minimise
  the contamination by foreground Galactic stars}; location slightly
beyond the boundary of the LG, so as to maximise the \evh{physical}
area sampled within the $1\arcmin$ \evh{field-of-view} (FoV) while
still detecting stars at the tip of the red giant branch (TRGB) with
an adequate S/N; existence of images of the same field in HST and/or
ESO archives; and the presence of a relatively strong intermediate age
component.

Currently, AO imagers operate only at near-IR wavelengths, which are
best suited to studying evolved stellar populations, in particular,
cool stars on the asymptotic giant branch (AGB).  This evolutionary
stage of low and intermediate mass stars is difficult to model because
of its sensitivity to uncertain input physics, like mass loss and
convection.  However, AGB stars provide a major contribution to the
integrated light of galaxies with intermediate-age stellar populations
\citep{renzbuzz1986}, therefore, it is very important to derive
information on the productivity of these stars.  This can be done by
analysing the stellar content of galaxies with a strong intermediate
age component, which is the motivation for our near-IR \evh{study} of
LG galaxies 
\citep{held+2007,gull+2007for,gull+2007sagdig}.

Ground-based optical photometry of \uks\ was first presented by
\cite{lee+1999}.  The colour-magnitude diagram (CMD) exhibits a
well-defined RGB, and a number of AGB stars. From the \evh{TRGB}
magnitude and from the colour of these stars, \cite{lee+1999} derive
the distance modulus and average metallicity of the galaxy as
$(m-M)_0=26.59\pm0.12$ and \feh$=-1.98$. More recently, from
photometry obtained with the WFPC2 on board of the HST,
\cite{kara+2002} measured $I^{\text{TRGB}}=22.72\pm0.12$, from which
they obtain $(m-M)_0=26.74\pm0.15$, corresponding to $2.23\pm 0.15$
Mpc. Thus, this galaxy is likely a member of the Sculptor Group. Since
the HST photometry is more accurate and since the distance
determination by \cite{kara+2002} is based on a more modern
calibration, in this paper, we will adopt the \cite{kara+2002} value.  This
implies that the absolute magnitude of the galaxy is $M_V = -13.24 $.
\referee{ Although \uks\ contains a young stellar component, there is
  no evidence of significant \ion{H}{II} emission
  \citep{mill1996,kais+2007}, which suggests a very low rate of
  ongoing star formation.  No mid-infrared emission from hot dust nor
polycyclic aromatic hydrocarbon is detected at 8\,${\mu}m$ \citep{jack+2006}
but it is embedded in a neutral hydrogen cloud that asymmetrically
covers the whole galaxy \citep[e.g.,][]{buyl+2006}.  The \ion{H}{I}
mass is $\sim 6 \times 10^6 M_\odot$, while for CO emission only an
upper limit on the molecular gas mass of $1.4 \times 10^5 M_\odot$ is
available \citep[][ans refs. therein]{buyl+2006}}.

\section{The data}\label{s:reduc}

\subsection{MAD observations }

MAD is a project \citep{marc+2007} mainly developed by ESO to test the
multi-conjugated adaptive optics (MCAO) capabilities on the sky in the
framework of the design of the European Extremely Large Telescope (ELT).
MAD was mounted on the UT-3 of the Very Large Telescope (VLT) to
realise the first MCAO observation on the sky \citep{bouy+2008}.
The instrument 
accommodates two wavefront sensors (WFS): a star-oriented
multi-Shack--Hartmann
and a
layer-oriented \citep[LO,][]{raga+2000,viar+2005} multi-pyramid
\citep{raga1996}.  
Both WFS use reference stars on a $2\arcmin$
technical FoV. 
MAD is complemented with the CAMCAO scientific IR camera, with a $2k\times
2k$ Hawaii\,{\sc ii} IR detector that can be moved across the
$2\arcmin$ corrected circular FoV. The pixel scale is $0\farcs028$
pixel$^{-1}$, yielding a $57\arcsec \times 57\arcsec$ square FoV on
the detector.

We took observations of \uks\ on Sept.~27, 2007 with the LO
wavefront sensor option, with the aim of testing single pyramid AO
observations in the bright-end regime. The reference star has
$V\approx 11.5$ and is located at $\sim 24 \arcsec$ from the centre of
the FoV.  This is the very first pyramid WFS AO-assisted science
observation \citep[see][for a discussion of the advantages of this
technique]{raga+1999} on an 8m-class telescope.  Results
from other observations with full multi-pyramid MCAO capabilities will
be presented elsewhere.

Our data set consists of 21 $J$ frames and 15 $K_s$ frames centred on
\uks, at $\alpha$(J2000)$= 23^h26^m27^s$,
$\delta$(J2000)$=-32\degr23\arcmin16\arcsec$.  The total integration
time is 37 and 30 min in $J$ and $K_s$ band, respectively.

\subsection{Reduction and photometric calibration}

We calibrated our near-IR photometry by
comparing 
stars in common with the 2MASS point-source catalogue
\citep{stru+2006} and by using 
$J$ and $K_s$ archive images obtained with the 
Son of ISAAC (SOFI)
camera mounted at
the ESO New Technology Telescope (NTT)
to define secondary photometric standards. 

We reduced both SOFI and MAD raw images following the standard
procedure for IR data, as described by \citet{gull+2007for}.
We paid careful attention to image alignment, allowing a correction for
possible field rotation.  We limited the area used in the final analysis
to a $45\arcsec\times45\arcsec$ region because of stray light
affecting one edge of the MAD images.
\referee{From the integration of the  $R$-band surface
  brightness profile \citep{lee+1999} of the galaxy, 
 we estimate that the luminosity fraction observed is $\sim 36$\%.}

The point spread function (PSF) of stellar objects on the combined MAD
images is fairly uniform across the whole frame, with deviations of
$\lesssim 10 \%$ of the full width at half maximum (FWHM). The mean FWHM measured on the $J$ an $K_s$
images is, respectively, $0\farcs15$ and $0\farcs11$.
This is a good result,
considering that \evh{the seeing was $0\farcs52$ and $0\farcs42$ in $J$
and $K_s$, respectively (estimated from the ESO DIMM Monitor
measurements in the $V$ band)} and that the diffraction limit for the
VLT is $0\farcs04$ for the $J$ band and $0\farcs07$ for the $K_s$ band.
The Strehl ratio measured on the $J$ and $K_s$ frames is 7.6\% and 21.4\%,
respectively.
The ellipticity of stellar images is small (9\% and 10\% in $J$ and $K_s$),
with an r.m.s. variation $\lesssim 5\%$ over the whole FoV.

We performed stellar photometry on MAD and SOFI images using
DAOPHOT/ALLSTAR programs \citep{stet1987}, adopting 
a Penny model for the PSF, with a quadratic dependence
on the position on the frame. 
The astrometric and photometric calibration of the SOFI data
was obtained using the USNO-A2.0 \citep{mone+1998}
and the 2MASS \citep{stru+2006} databases.
We then used the resulting catalogue as a reference for the
astrometric and photometric calibration of MAD frames. 
Considering the uncertainty of our
two-steps calibration, the final error on the zero-point of
MAD photometry resulted $\simeq$ 0.15 mag. 

Finally, the near-IR data were complemented with optical WFPC2/HST
data from \cite{holt+2006}. This allows us to test the spatial resolution of 
MAD images, as well as take advantage of a wide colour baseline to study
the stellar population.

\begin{figure*}
  \centering
  \includegraphics[width=.30\textwidth]{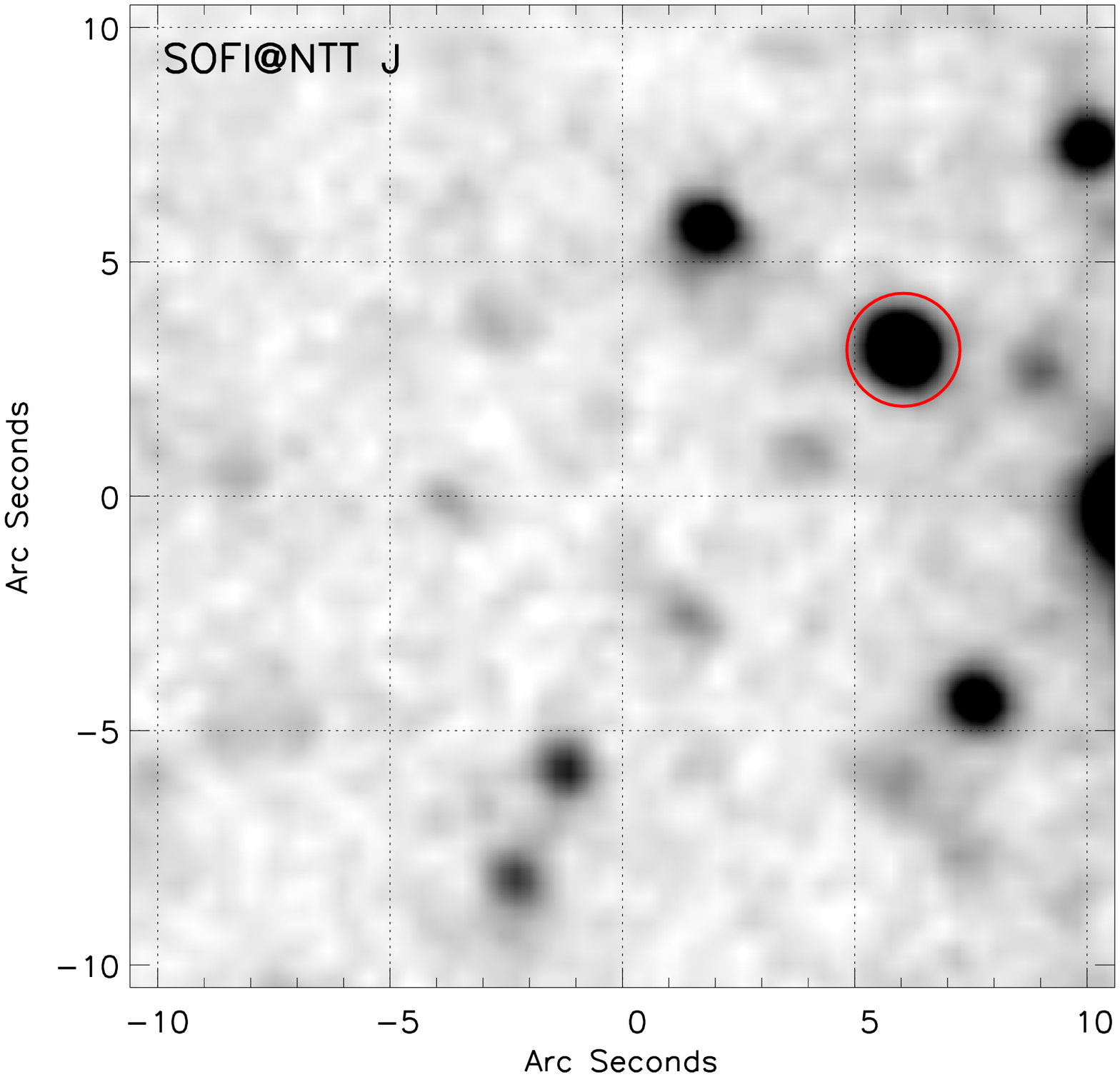}
  \includegraphics[width=.30\textwidth]{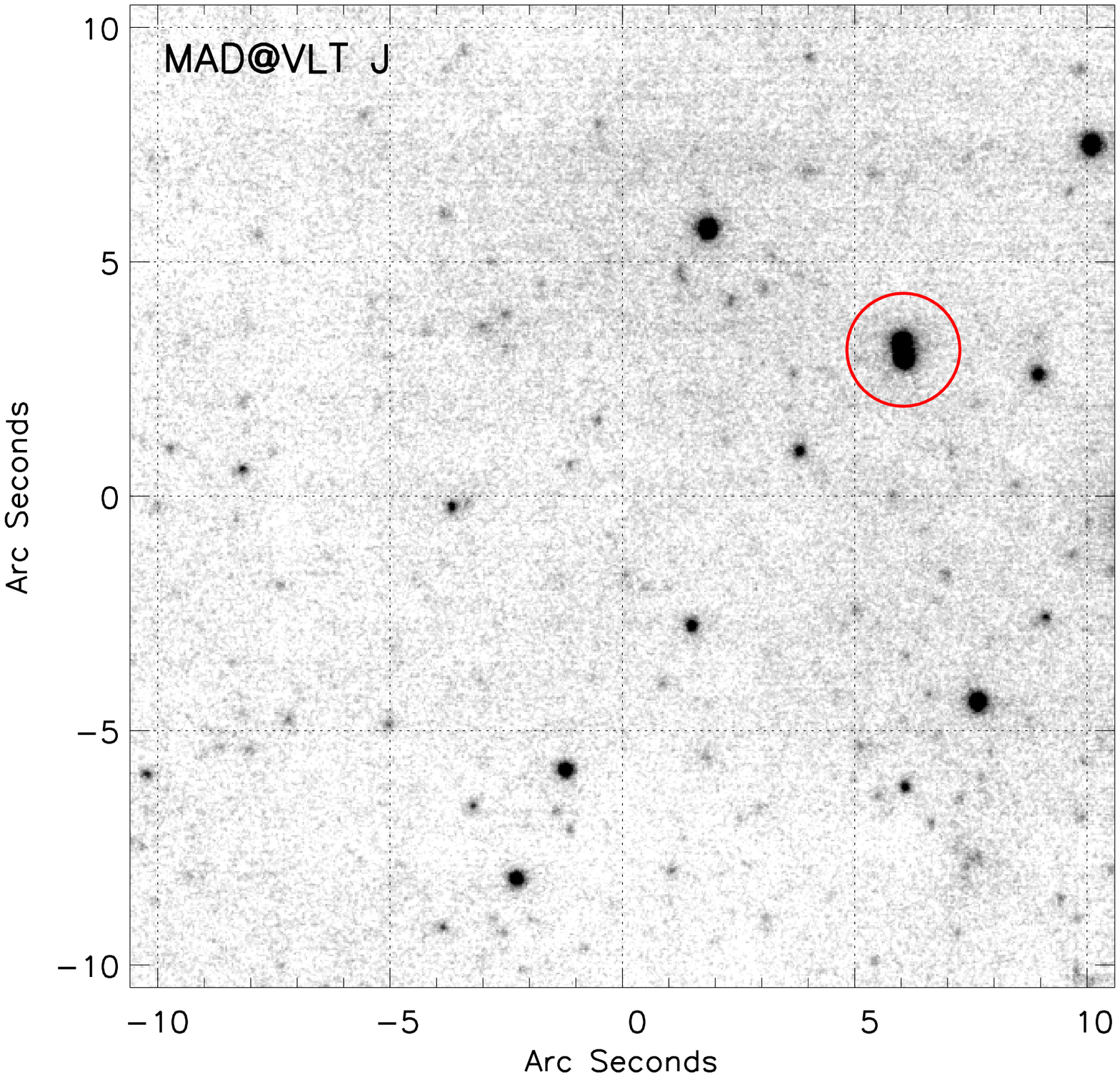}
  \includegraphics[width=.30\textwidth]{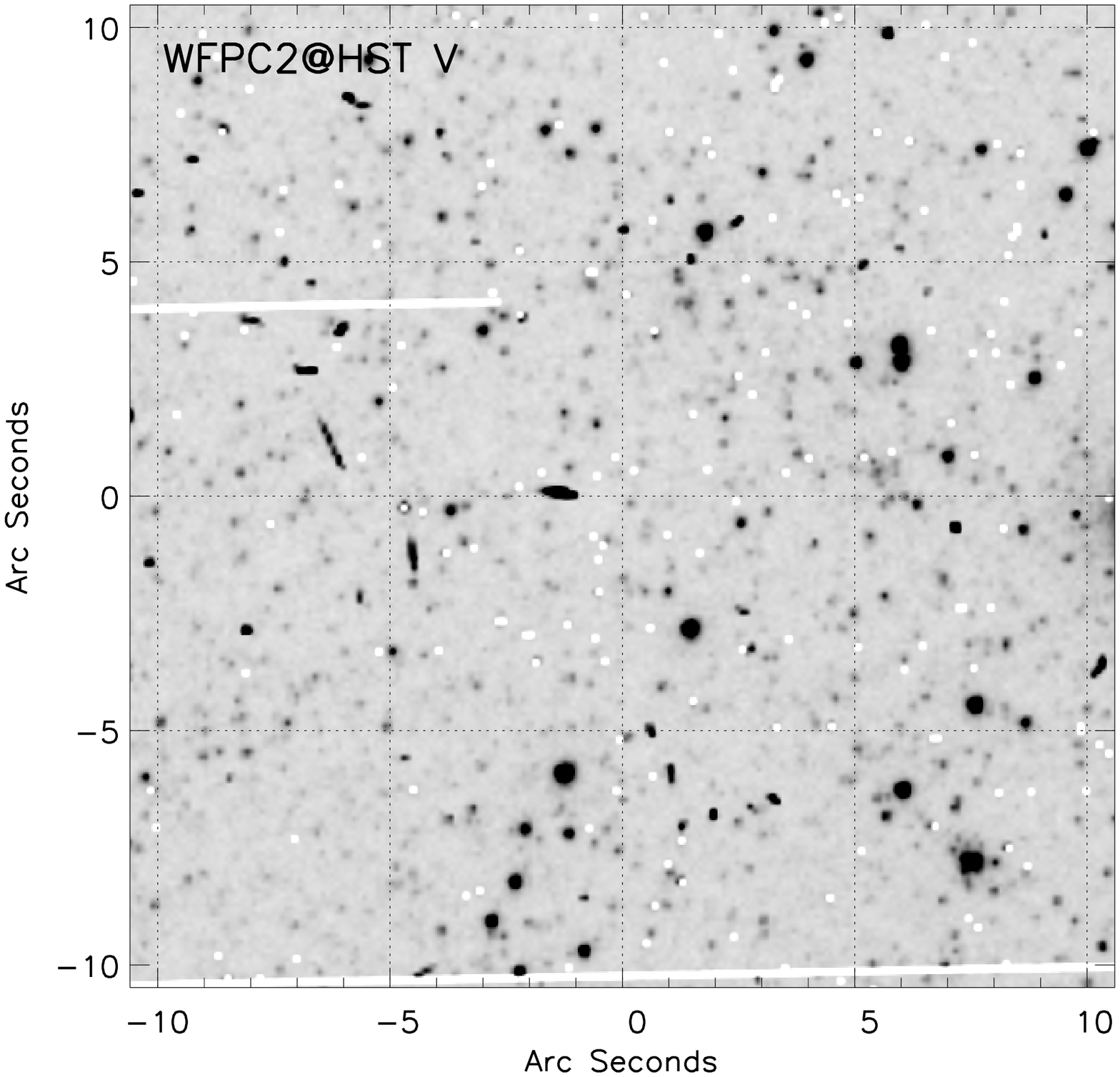}
   \caption{
SOFI, MAD and WFPC2 images of the same region in \uks.
Only a section of the full area analysed in our study is shown
to better illustrate the details. 
The two circles 
highlight the effect of the higher spatial resolution of MAD versus SOFI. 
}
\label{f:maps}
\end{figure*}

Figure \ref{f:maps} \evh{compares} the 3 images from SOFI, MAD, and WFPC2
for the same region in \uks. The improvement in resolution 
between MAD and SOFI image is clearly apparent. 
This implies a significantly better photometry  of faint objects 
and, in particular, the possibility of obtaining accurate photometry for faint
stars that are embedded in the halo of brighter stars.

In order to evaluate the completeness and photometric errors of our
catalogue, we performed an extensive set of 160 artificial star
experiments using $\sim$1000 stars for each run.  Input magnitudes
were randomly generated to reproduce an uniform distribution over the
colour and magnitude range of real stars in our image ($0<J-K_s<2$ and
$15<K_s<22.5$).
We found that our photometry is complete at the 50\% level down to $K_S\simeq20.7$.
At this magnitude level, the photometric error is $\sim0.1$ mag in the
$K_s$ band, and a factor of 2 lower in the $J$ band.
These results are illustrated in Fig.~\ref{f:cmd_mad_sofi}.

\section{The evolved stellar populations of \uks}

\begin{figure}
\resizebox{\hsize}{!}{\includegraphics{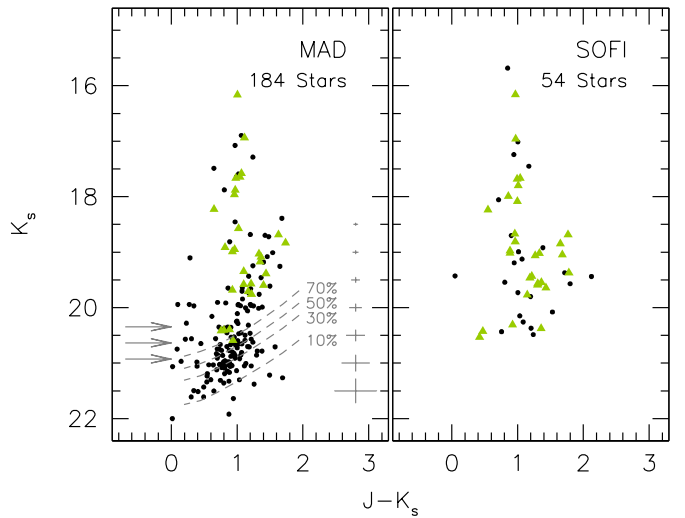}}
  \caption{CMD obtained from MAD data ({\it left}) and from SOFI data
    ({\it right}) for the same FoV. \referee{In the left panel, the
      completeness levels ({\it dashed lines}) and photometric errors
      ({\it crosses}) from artificial star experiments are shown.}  The
    expected location of the TRGB is indicated with arrows for
    $\feh=-1.0$, $-1.5$, and $-2.0$ (top to bottom).  \referee{Stars
      used as secondary photometric standards are shown as triangles.}}
\label{f:cmd_mad_sofi}
\end{figure}

The CMD of \uks\ obtained from MAD images is shown  in
Fig.~\ref{f:cmd_mad_sofi} and compared to that obtained from SOFI
data in the same FoV.
The MAD CMD is about 1 magnitude deeper; more
importantly, the photometric accuracy is higher, leading to a
much better definition of the sequences on the CMD. This is mostly due
to the higher spatial resolution of MAD, which allows us to resolve stars
that are blended on SOFI images.

The red tail of bright stars in the MAD CMD, extending up to
$J-K_S\simeq1.8$ and $K_s\simeq18.5$, is consistent with the locus
of carbon-rich AGB stars \citep[c.f. other recent
near-IR studies, e.g.,][]{gull+2007for,gull+2007sagdig,menzies+2008}.
It is tempting to locate the TRGB at $K_s\simeq 20.5$, where a
discontinuity is apparent in the stellar magnitude distribution.
However, \evh{a formal measurement} of the TRGB from the luminosity
function cannot be derived because of the incompleteness of our
photometry at these magnitudes. Since the distance modulus is known
from optical observations, we can estimate the expected level of the
TRGB: according to \cite{vale+2004} calibration, the absolute $K_s$
magnitude of the TRGB depends on metallicity, being $-5.82$, $-6.11$, and
$-6.40$ for $\feh=-2,-1.5$, and $-1$, respectively.  Assuming 
a reddening $\ebv=0.015$ 
\citep{schl+1998}, we get $K_s^\text{TRGB}=20.93,20.64$, and $20.35$,
as the metallicity increases. These three values are indicated with
arrows in Fig.~\ref{f:cmd_mad_sofi}: for the metallicity determined by
\cite{lee+1999} ($\feh \simeq -2$), the TRGB should be close to the
limit of our photometry. 
The \citet{lee+1999}
determination is based on a relatively shallow CMD. Comparing the
deeper $(V,I)$ HST CMD by \cite{holt+2006} to the fiducial lines of
Galactic globular clusters by \cite{dacoarma1990}, 
we obtain a mean metallicity
$\feh \simeq -1.7$ for the RGB stars in this galaxy.
This has to be regarded as a lower limit, since the bulk of \uks\
stellar population is younger than Galactic globular clusters.  As an
example, \cite{savi+2000} estimated that for a $\sim5$ Gyr stellar
population, the age correction to be applied to the metallicity obtained
with our method is $+0.4$ dex. The location of the TRGB would then come
close to the discontinuity of the star's distribution mentioned above.
In the following, we adopt $\feh = -1.5$, which yields a TRGB magnitude
of $K_s=20.65$; we verified that this metallicity is compatible with the
mean $V-K_s$ colour of the RGB in \uks.

\realfigure{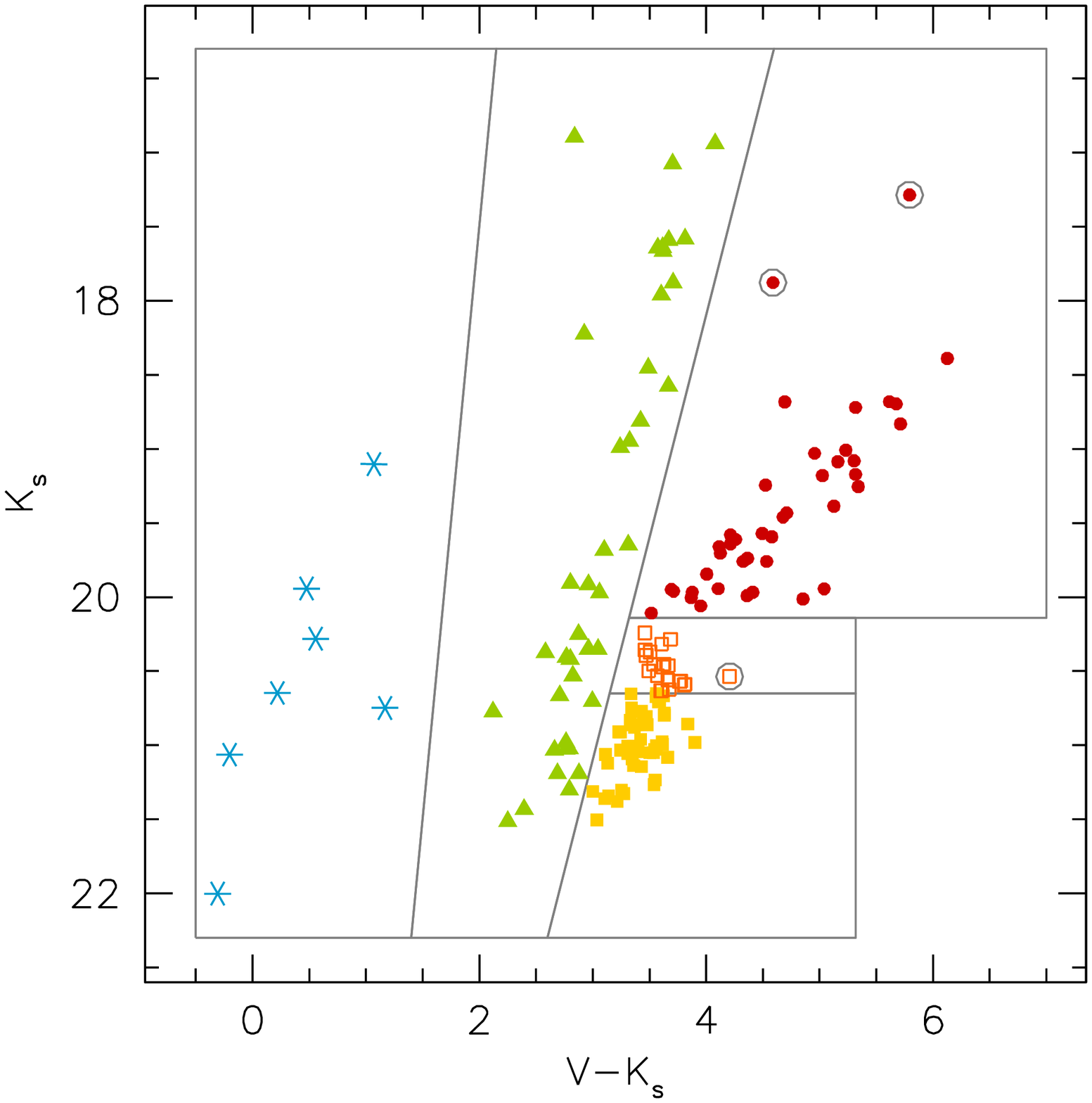}{Combined optical--near-IR MAD CMD of \uks.
We used this CMD to select blue supergiants ({\it starred symbols}),
red supergiants ({\it triangles}), RGB ({\it squares}), and AGB
stars ({\it filled
circles}).  \evh{{\it Open squares}} are the stars with an
uncertain classification, but likely E-AGB stars (see text for details).
The three stars marked by {\it open circles} have peculiar colours
and are possibly photometric blends.  }{f:cmdvkk}

Figure \ref{f:cmdvkk} shows the $V-K_s$ {\it vs} $K_s$ CMD obtained by
combining the HST and MAD photometry. This colour combination is
particularly well suited to distinguish the different evolutionary
sequences.
\referee{
The contamination by foreground stars in our fields is negligible. 
In fact, using simulations of  the Milky Way population 
performed with  the TRILEGAL code \citep{gira+2005}, we expect only  
  three foreground stars in the magnitude and colour range of our CMD.
    }
Guided by the optical CMD, on this figure we draw the
lines bordering the areas occupied by blue supergiants, red
supergiants, AGB, and RGB stars. The blue and red supergiants
are core Helium burning stars with masses down to $\sim 5 M_\odot$,
resulting from the star formation activity occurred over the last
$\sim 100$ Myr. 
\referee{ The main-sequence progenitors of this young population are
  sampled in the optical CMD by \cite{holt+2006}, in which the blue
  plume contains stars with ages from $\simeq 80$ to $\simeq 10$ Myr
  old. A quantitative interpretation of this component needs
  detailed simulations that take into account photometric errors and
  completeness of the HST data.  Here we concentrate on the intermediate age
  component for which the near-IR CMD offers a better diagnostic than the
  optical diagram.}
The bright portion of the red sequence, extending from $K_s \simeq
20.1$ up to $K_s=18.5$ and with very red colours (up to $V-K_s=6.5$) hosts
bright AGB, mostly carbon (C) stars, while below the TRGB (at
$K_s=20.65$) we sample the oldest stars. The stars in
the intermediate region ({\it open squares}, $20.65>K_s>20.1$)
\referee{are probably AGB stars, but some of them could actually be
  high-metallicity RGB stars. The uncertainty stems from the
  dependence of the TRGB $K_S$-magnitude on the metallicity discussed
  above, and on  these stars being located
 (on the optical CMD)  just below the TRGB in the $I$ band.}

\referee{Our  observations can be used to
derive a rough estimate of the star formation that occurred in \uks\
at intermediate ages by considering the number of AGB stars brighter
than the TRGB, which is proportional
to the gas mass converted into stars between $\sim 0.1$ and a few
Gyrs \citep[e.g.,][]{greg2002}.
The {\it specific production} of bright AGB stars
(i.e., $\delta n_{\rm AGB,b}$, the number of stars per unit mass
of the parent stellar population that fall in this region of the CMD)
depends on age and metallicity.
We have determined $\delta n_{\rm{AGB,b}}$ as a function of age, for a
sample of globular clusters in the 
Large Magellanic Cloud
for which near-IR CMDs, ages,
and total luminosities are known from the literature. From these,
 we  found that
it reaches a maximum ($\sim 2 \times 10^{-4} M_\odot^{-1}$) at ages $\sim 1$
Gyr, to drop significantly at older ages, down to $\sim 3 \times 10^{-5}
M_\odot^{-1}$ at $\sim$ 3 Gyr (close to the limit of our calibration).
The stellar distribution in our CMD is suggestive of an extended
episode of star formation; by averaging our empirical calibration, we
obtain a specific production of 0.15 or 0.08 stars per $10^3 M_\odot$,
if this episode started 1.5 or 3 Gyr ago, respectively. Since we count
59 objects on the red sequence at magnitudes brighter than the TRGB,
we bracket the stellar mass produced at intermediate ages in the range
between 4 and 7.5 $\times 10^{5} M_\odot$. We note that these estimates are based on a straight
Salpeter initial mass function (IMF). If a \cite{chab2005} IMF were
assumed, the masses would be $\sim 0.65$ times smaller.}

\realfigure{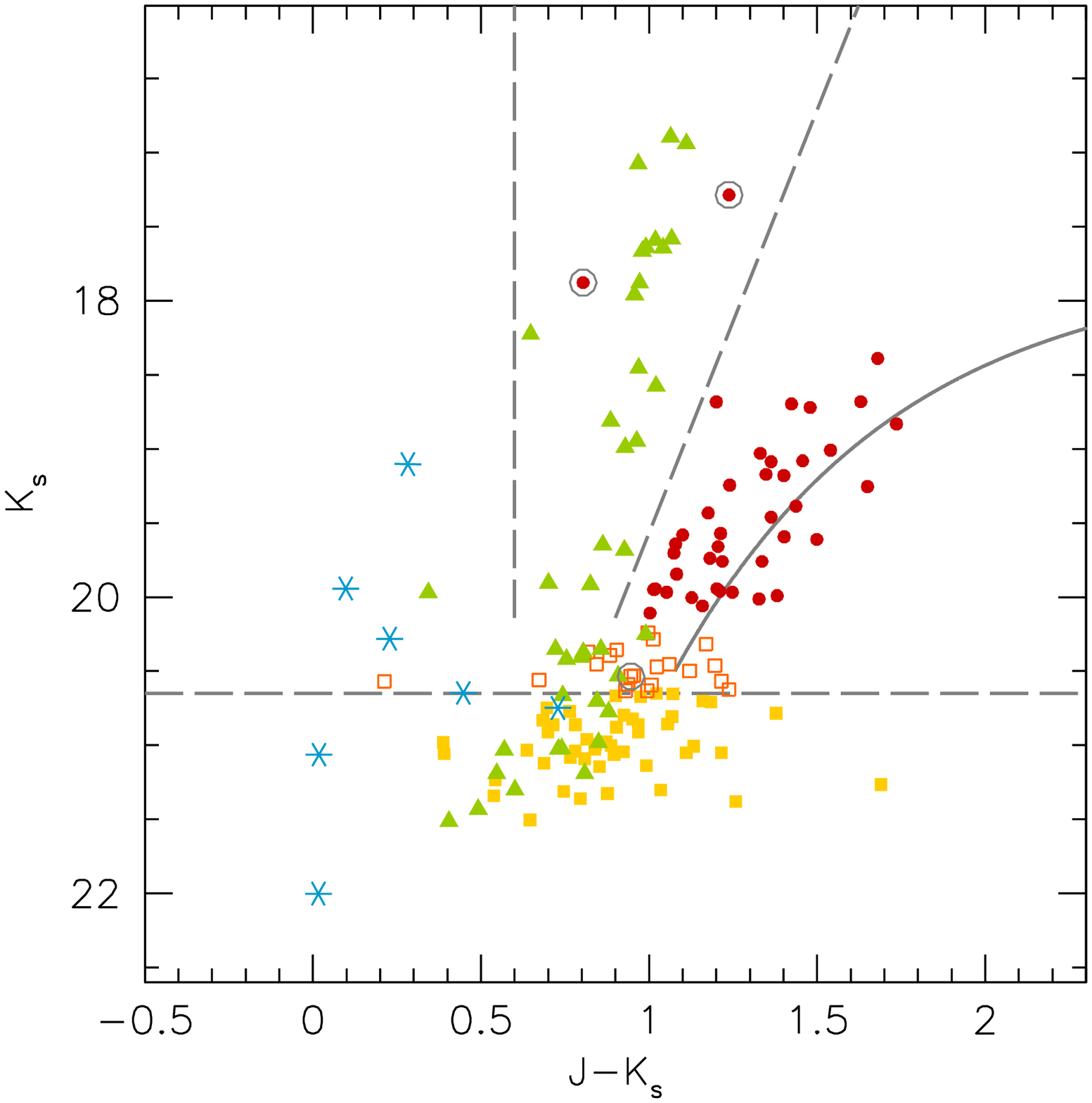}{ The near-IR CMD of \uks.  The different
  symbols refer to the selection presented in Fig.~\ref{f:cmdvkk}.  The
  solid line is the main locus of C-star in nearby dwarf galaxies
  defined by \cite{tott+2000} \evh{shifted to the adopted distance of
  \uks}.}{f:cmdjkk}

Using the classification obtained from the ($V,K_s$) CMD, we construct the
($J,K_s$) diagram to see how the various
sequences can be identified when only IR data are available
(see Fig.~\ref{f:cmdjkk}). Although the
sequences here are less well traced, the various evolutionary stages are relatively well
separated. In particular, this CMD shows 
that all stars in the red tail are located in
the position expected for C-stars, since they are compatible with
the main locus of C-stars in nearby dwarf galaxies defined by
\cite{tott+2000}.
\referee{
These results confirm that near-IR CMDs are a very powerful tool
to clearly detect C-stars, as indicated by our previous studies of LG galaxies 
\citep{gull+2007for,gull+2007sagdig}.
}

To summarise our work, we obtained a complete and accurate census of
the bright evolved stellar population (red supergiants and AGB stars) in
\uks, a dwarf irregular galaxy at a distance of 2.23 Mpc.
We have shown that with near-IR AO images at 8m class telescopes
it is possible to investigate the SFH in galaxies well beyond the LG.  
Considering the technical limitation of the {\it demonstrator}, we believe that
these results forecast very promising opportunities for this kind
of studies with advanced AO at ELT.

\bibliographystyle{aa} 
\bibliography{bibliouks} 
\end{document}